\documentclass[fleqn,twoside]{article}
\usepackage{espcrc2}
\usepackage{epsf}
\makeatletter
\newcommand\nameaddress[1]{{\addtocounter{address}\m@ne
                            \expandafter\xdef
                               \csname address.#1\endcsname
                               {\number\value{address}}%
                            \addtocounter{address}\@ne}}
\newcommand\useaddress[1]{{\edef\doit{\noexpand\addressmark
                                      \noexpand\setcounter{address}%
                                      {\number\value{address}}}%
                           \setcounter{address}{\csname address.#1\endcsname
                                                }%
                           \expandafter}\doit}
\newcommand\anotheraddress[1]{{\let\orig@makeadmark=\@makeadmark
                               \def\@makeadmark##1{\orig@makeadmark
                                                   {\negthinspace,##1}}%
                               \address{#1}}}
\newcommand\slashnext[1]{\mathpalette{\bgroup\let\style=}
                                     {\setbox0=\hbox{$\style #1$}%
                                      \setbox2=\hbox to\wd0{\hss$\style/$\hss}%
                                      \hbox to 0pt{\box2\hss}\box0\egroup}}
\@ifundefined{inlinecite}{\let\inlinecite=\cite}
\makeatother
\begin{document}
\title{\begin{flushright}\normalsize
            LA-UR-99-1717 \\
	    UW-PT/98-15 \\
	    UTCCP-P-56 \\
	    DUKE-TH-99-185
       \end{flushright}
       Non-perturbative Renormalization Constants using Ward Identities}
\author{Tanmoy Bhattacharya,\thinspace
        \address{MS B-285, Los Alamos National Lab, Los Alamos,
		 New Mexico 87545, USA}%
        \nameaddress{lanl}
	Shailesh Chandrasekharan,\thinspace
	\address{Department of Physics, Duke University, Durham,
        	 North Carolina 27705, USA}
	Rajan Gupta,\thinspace
        \useaddress{lanl}
	Weonjong Lee,\thinspace
        \useaddress{lanl}
        and Stephen Sharpe\thinspace
	\address{Physics Department, University of Washington,
          	 Seattle, Washington 98195, USA}%
	\anotheraddress{Center for Computational Physics, University
                 of Tsukuba, Tsukuba, Ibaraki, 305-8577, Japan}
	}

\begin{abstract}

We extend the application of axial Ward identities to
calculate $b_A$, $b_P$ and $b_T$, coefficients that
give the mass dependence of the renormalization constants of 
the corresponding bilinear operators in the quenched theory. 
The extension relies on using operators with non-degenerate quark masses. 
It allows a complete determination of the $O(a)$ improvement coefficients
for bilinears in the quenched approximation using Ward Identities alone.
Only the scale dependent normalization constants $Z_P^0$ (or $Z_S^0$)
and $Z_T$ are undetermined.
We present results of a pilot numerical study using hadronic correlators.

\end{abstract}
\maketitle

\nopagebreak

Wilson's discretization of the Dirac operator introduces lattice
artifacts at $O(a)$, the effects of which
are large for lattice spacings accessible in present simulations.
It is thus expedient to devise improved lattice discretizations,
and significant recent progress has been made in this direction.
In particular the ALPHA collaboration has implemented
Symanzik's improvement program~\cite{symanzik},
in which both the action and external sources are improved
by the addition of higher dimensional operators.
A key ingredient is the development of methods to determine the
coefficients of these extra operators, the ``improvement coefficients'',
non-perturbatively. In this note we present an extension of
these methods which allows the determination of all improvement
coefficients for bilinear operators.

We begin by reviewing the results of the ALPHA collaboration.
They have shown that $O(a)$ artifacts can be removed from
on-shell quantities by the addition of a single local operator of dimension
five~\cite{alpha-0}, resulting in the Sheikholeslami-Wohlert (or ``clover'')
action~\cite{sw-1}
\begin{eqnarray}
	S &=& S_{G} + S_{W} + S_{SW} 	\\
	S_{SW} &=&  a^5 c_{SW}\ \sum_{x} \bar{\psi}(x) \frac{i}{4}
	\sigma_{\mu\nu} F_{\mu\nu}(x) \psi(x) \,.
\end{eqnarray}
Here $ S_G $ and $S_W$ are respectively Wilson's gluon and quark actions, 
and $S_{SW}$ is the Sheikholeslami-Wohlert term 
($ F_{\mu\nu} $ is the lattice gluon field strength tensor).
Improvement of quark bilinears is slightly more involved~\cite{alpha-0}.
We consider only flavor off-diagonal operators,
the bare lattice forms of which are
\begin{eqnarray}
A^{(jk)}_{\mu}    &\equiv& \bar{\psi^j} \gamma_{\mu} \gamma_{5} \psi^k \,,\\
 V^{(jk)}_{\mu}   &\equiv& \bar{\psi^j} \gamma_{\mu}  \psi^k \,, \\
 P^{(jk)}         &\equiv& \bar{\psi^j} \gamma_{5}  \psi^k \,, \\ 
 S^{(jk)}         &\equiv& \bar{\psi^j}  \psi^k \,, \\
 T^{(jk)}_{\mu\nu}&\equiv& \bar{\psi^j} i \sigma_{\mu\nu}  \psi^k \,,
\end{eqnarray}
with $j\ne k$ being flavor indices, $\sigma_{\mu\nu} =
[\gamma_\mu,\gamma_\nu]/2$, $\gamma_5 = \gamma_1 \gamma_2 \gamma_3
\gamma_4$, and the hermitian $\gamma$ matrices satisfy $\{\gamma_\mu,
\gamma_\nu\} = 2\delta_{\mu\nu}$.  Removal of $O(a)$ errors from the
on-shell matrix elements of these bilinears requires both the addition
of extra operators (except for $P$ and $S$),
\begin{eqnarray}
(A_I)_{\mu}    & \equiv & A_{\mu} + a c_A \partial_\mu P \\
(V_I)_{\mu}    & \equiv & V_{\mu} + a c_V \partial_\nu T_{\mu\nu} \\
(T_I)_{\mu\nu} & \equiv & T_{\mu\nu} +
                a c_T ( \partial_\mu V_\nu - \partial_\nu V_\mu) \,.
\label{eq:impbilinears}
\end{eqnarray}
and the introduction of mass dependence,
\begin{eqnarray}
(X_R)^{(ij)}    & \equiv & Z_X^0(1+b_X am_{ij} ) (X_I)^{(ij)} \,.
\label{def_XA} 
\end{eqnarray}
Here $X= A, V, P, S, T$,
the $Z_X^0$ are renormalization constants in the chiral limit, 
and  $m_{ij} \equiv ( m_i + m_j)/2$ is the average bare quark mass.

Complete $O(a)$ improvement of matrix elements requires
that the coefficients $c_{SW}$,  $c_X$, and $b_X$, 
as well as the matching constants $Z_X^0$,
be determined non-perturbatively.\footnote{%
For brevity we refer to the $Z_X^0$ as improvement coefficients in the
following.}
Previous work has shown how the enforcement of
axial and vector Ward identities (WI) allows one to determine
$Z_V^0$, $Z_A^0$ and $Z_P^0/Z_S^0$ \cite{rome-0}, 
$c_{SW}$, $c_A$ and $b_V$ \cite{alpha-1,alpha-2,alpha-3}, 
$c_V$ \cite{alpha-4}, $c_T$ \cite{rome-1},
and $b_P - b_A$ and $b_S$ \cite{rome-3}.
We discuss here an extension that yields $b_A$, $b_P$, and $b_T$.\footnote{%
A preliminary account of this work was given in Ref.~\cite{lat98us-1}.}
This provides an alternative to the
non-perturbative method proposed in Ref.~\cite{rome-1}
which uses the short-distance behavior of two point functions.
The two remaining constants $Z_P^0$ (or $Z_S^0$) and $Z_T^0$ are scale
and scheme dependent, and cannot be determined using WI.  

It is important to note that
the relations we derive do not extend directly to the unquenched
theory, which requires additional improvement constants and a more
complicated set of conditions as will be presented
in~\inlinecite{WIunquenched}.

%
%
We begin by recalling the ALPHA method for determining
$c_{SW}$ and $c_A$ \cite{alpha-2}. 
The improved axial current should satisfy
\begin{equation}
\partial_\mu (A_R)^{(ij)}_{\mu} =  (m^R_i+m^R_j)
(P_R)^{(ij)} + {\cal O}(a^2)\,,
\end{equation}
when inserted between on-shell states.  Here $m^R$ is the renormalized
quark mass.  It follows that the ratio
\begin{eqnarray}
2 \tilde m_{ij} &\equiv&  \frac{ \sum_{\vec{x}} \langle 
 \partial_\mu [A_\mu + 
  a c_A \partial_\mu P]^{(ij)}(\vec{x},t) J^{(ji)}(0) \rangle} 
 {\sum_{\vec{x}} \langle P^{(ij)}(\vec{x},t) J^{(ji)}(0) \rangle} 
\label{cA} \\
&=&
(m^R_i+m^R_j)  \frac{Z_P^0(1+b_P am_{ij})}{Z_A^0(1+b_A a m_{ij})} 
  \,.
\label{m(t)}
\end{eqnarray}
should be independent both of the choice of sources $J$
and of the time $t$, as long as $t\ne 0$, 
up to corrections of $O(a^2)$.
This is to be achieved by simultaneously tuning $ c_{SW} $ and $c_A$. 
Our implementation of this condition
differs from the Schr\"odinger functional method of
Ref.~\cite{alpha-2} in that we use standard two-point correlation
functions, with a variety of choices of sources.
We also fix $c_{SW}$ {\em a priori} and use the condition to determine
only $c_A$.

It is convenient for the following discussion to introduce $b$
coefficients defined using the masses $\tilde m_{ij}$, which we refer
to as WI masses, rather than the bare quark mass, i.e.
\begin{equation}
(X_R)^{(ij)}    \equiv Z_X^0(1+\tilde b_X a \tilde m_{ij} ) (X_I)^{(ij)} \,.
\label{def_tildeb} 
\end{equation}
The advantage of the WI mass is that it can be determined,
using eq.~(\ref{cA}), without the need for chiral extrapolation.
By contrast, to determine the bare quark mass,
 $a m=1/(2\kappa)-1/(2\kappa_c)$,
one needs to know the critical hopping parameter, $\kappa_c$,
the calculation of which requires chiral extrapolation.
The difference is particularly significant when using standard two-point
functions as opposed to the Schr\"odinger functional.
To determine the standard $b_X$ from the $\tilde b_X$ we need the
relation between the bare and WI masses, which is given in 
eq.~(\ref{mass2}) below. In the quenched theory it follows that
\begin{equation}
b_X = {Z_P^0 \over Z_A^0 Z_S^0}\tilde b_X  + O(a) \,.
\label{brelation}
\end{equation}
Note that we only need this relation at leading order in $a$, 
since the $b_X$ appear only in $O(a)$ corrections.

%
%
With $c_{SW}$ fixed,
{\bf $ Z_V^0$} and {\bf $\tilde b_V$} can be obtained in
the standard way using charge conservation.
We use the forward matrix elements of $(V_I)_4$ between pseudoscalars,
\begin{eqnarray}
& &  \frac{1}{ Z_V^0 (1+\tilde b_V a\tilde m_2) } = \nonumber \\
& & \hskip 0.3in   \frac{ \sum_{\vec{x}, \vec{y}}
  \langle P^{(12)}(\vec{x},\tau) 
	(V_I)_4^{(22)}(\vec{y},t) J^{(21)}(0) \rangle }
  { \langle \sum_{ \vec{x}} P^{(12)}(\vec{x},\tau) J^{(21)}(0) \rangle } \,.
\label{ZV}
\end{eqnarray}
with $ \tau > t > 0 $ and $ J = P $ or $A_4$.
Here and below we use an abbreviation for the WI mass for two degenerate
flavors: $\tilde m_i =\tilde m_{ii}$.
Note that the $c_V$ term in $V_I$ does not contribute to the r.h.s.,
and so is not determined by this condition.

The remaining improvement constants are determined by enforcing the generic
form of the integrated axial WI (AWI), in which the operator 
$ {\cal O}^{23} = {\bar\psi}^{(2)} \Gamma \psi^{(3)} $
is transformed into 
$ \delta {\cal O}^{13} ={\bar\psi}^{(1)} \gamma_5 \Gamma \psi^{(3)} $
by a chiral rotation in the $1,2$ flavor subspace:
\begin{equation}
\biggl\langle 
	  \delta {\cal S}^{(12)} {\cal O}_R^{(23)}(y) J^{(31)}(z) \biggl\rangle
    = \biggl\langle \delta {\cal O}_R^{(13)}(y) \  J^{(31)}(z)\biggl\rangle 
	\,.
\label{AWI} 
\end{equation}
%
%
Here $\delta{\cal S}$ is the variation in the action
\begin{equation}
 \delta {\cal S}^{(12)} = \int_{\cal V} d^4x 
        \biggl[ (m^R_1+m^R_2) (P_R)^{(12)} - 
\partial_\mu (A_R)^{(12)}_\mu \biggl]  \,,
\label{deltaS12}
\end{equation}
and the chiral rotation is restricted to a 4-volume ${\cal V}$ 
containing $y$ but not $z$.
The Ward identities should hold up to corrections of $O(a^2)$
if the operators and action are appropriately improved.
There is, however, an obstacle to implementing these constraints,
arising from the integral over $x$ in $\delta{\cal S}$.
This brings the pseudoscalar density $P_R(x)$ into contact with 
the operator ${\cal O}_R(y)$ on the l.h.s. of~(\ref{AWI}),
implying that on-shell improvement of $P$ and ${\cal O}$
is insufficient to improve the AWI.
(The integral over $\partial_\mu A_\mu$ gives a surface term
which does not involve contact with ${\cal O}$.)
The problematic term is explicitly proportional to quark masses,
and thus is absent in the chiral limit.
For this reason the AWI has previously been used only in the chiral limit,
from which one can determine $Z_A^0$, $Z_P^0/Z_S^0$, $c_V$ and $c_T$,
but not the $\tilde b_X$.

Our new observation is that by looking at the detailed dependence
on the quark masses one can work away from the chiral limit
and yet avoid contact terms. 
This allows the determination of certain
linear combinations of the $\tilde b_X$.
The cost is the need to use non-degenerate quarks.

For all but the coefficient $\tilde b_T$, it is sufficient to work
with two non-degenerate quarks: $m_1=m_2 \ne m_3$.  Since the contact
term is explicitly proportional to $m_1^R+m_2^R$, which itself is
proportional to $\tilde m_{12}$ at small quark masses, one can remove
it by extrapolating to $\tilde m_1=\tilde m_2=0$. The remaining
$\tilde m_3$ dependence allows one to determine combinations of the
$\tilde b_X$. In the following, we describe how this works for the
different bilinears, and explain our particular implementation.  The
extrapolation to $\tilde m_1=\tilde m_2=0$ will be implicit throughout
this discussion.\footnote{ In practice we keep the term $\int
(m_1+m_2) (P_R)^{(12)}$ in eq. (\ref{AWI}) prior to extrapolation,
since this increases the range of $t_{\cal O}$ for which the ratio of
correlation functions is constant.}

%
%
As a first application we show how to obtain $\tilde b_A-\tilde b_V$,
as well as $c_V$, using the AWI with ${\cal O}=V_\mu$.
The identity for the time component can be written 
\begin{eqnarray}
 r_1 &\equiv&  
\frac	{ \sum_{\vec{y}}
	\langle \delta {\cal S}^{(12)}_I 
	\ (V_I)_4^{(23)}(\vec{y},y_4) \  J^{(31)}(0) \rangle }
	{ \sum_{\vec{y}} 
	\langle (A_I)_4^{(13)}(\vec{y},y_4) \ J^{(31)}(0) \rangle }
\nonumber \\
   &=&  \frac{ Z_A^0 (1+ \tilde b_A a \tilde m_3/2) } 
{ Z_A^0 \cdot Z_V^0 (1+ \tilde b_V a \tilde m_3/2) }\,,
\label{cV1} 
\end{eqnarray}
where $\delta {\cal S}_I = - \int_{\cal V} \partial_\mu (A_I)_\mu$.
For the source we take $J=P$ or $A_4$. 
We have chosen to put the current at $\vec p=0$
so that the $c_V$ term in $V_I$ does not contribute.
Since we know $c_{SW}$ and $c_A$ we can calculate $r_1$,
and determine $\tilde b_A- \tilde b_V$ from its dependence on $\tilde m_3$.
The intercept provides a determination of $Z_V^0$ independent of
that from eq.~(\ref{ZV}).
The AWI for the spatial components can be written as
\begin{eqnarray}
\tilde r_1 \hspace{-7pt}&\hspace{-7pt}=\hspace{-7pt}&\hspace{-7pt}
\frac{ \sum_{\vec{y}} e^{i\vec{p} \cdot \vec{y} } 
        \langle \delta {\cal S}^{(12)}_I \ 
	[ V_i + a c_V \partial_\mu T_{i\mu}]^{(23)}(\vec{y},y_4) \  
	A_i^{(31)}(0) \rangle }
        { \sum_{\vec{y}} e^{i\vec{p} \cdot \vec{y} }
		\langle (A_I)_i^{(13)}(\vec{y},y_4) \  A_i^{(31)}(0) \rangle }
\,,\nonumber\\
    &=& \ r_1 \,.
\label{cV2} 
\end{eqnarray}
Enforcing this equality (for any small $\tilde m_3$) provides
a determination of $c_V$.

%
%
Reversing the roles of vector and axial bilinears provides a second
determination of $\tilde b_A-\tilde b_V$. For example, 
if $c_V$ is known from above, then one can use the ratio
\begin{eqnarray}
r_2 &\equiv &
\frac	{ \sum_{\vec{y}} e^{i\vec{p} \cdot \vec{y} }
	\langle \delta {\cal S}^{(12)}_I 
	    \ (A_I)_i^{(23)}(\vec{y},y_4) \  V_i^{(31)}(0) \rangle }
	{ \sum_{\vec{y}}  e^{i\vec{p} \cdot \vec{y} }
	\langle (V_I)_i^{(13)}(\vec{y},y_4) \ V_i^{(31)}(0) \rangle }
\nonumber \\
 &=&  \frac{ Z_V^0 (1+\tilde b_V a \tilde m_3/2) } 
     { Z_A^0 \cdot Z_A^0 (1+ \tilde b_A a \tilde m_3/2) }\,.
\label{ZAZV-1}
\end{eqnarray}
Given $Z_V^0$, this also yields $Z_A^0$.
The same information can be obtained from the combinations
\begin{eqnarray}
	& & \frac{1}{ \sqrt{ r_1 \cdot r_2} } = Z_A^0 \,,
\label{ZAZV-2}
\\
	& & \sqrt{ \frac{r_1} {r_2} } =
	\frac{Z_A^0}{Z_V^0} (1+ (\tilde b_A - \tilde b_V) a\tilde m_3/2) \,.
\label{ZAZV-3}
\end{eqnarray}

%
%
Applying the method to the AWI with ${\cal O}=P$ and $S$ gives
two determinations of $\tilde b_P- \tilde b_S$. 
For example, one can use the $\tilde m_3$ dependence of
the ratio
\begin{eqnarray}
r_3 &\equiv &
\frac	{ \sum_{\vec{y}} e^{i\vec{p} \cdot \vec{y} }
	\langle \delta {\cal S}^{(12)}_I 
		\ S^{(23)}(\vec{y},y_4) \ J^{(31)}(0) \rangle }
	{ \sum_{\vec{y}}  e^{i\vec{p} \cdot \vec{y} }
	\langle P^{(13)}(\vec{y},y_4) \ J^{(31)}(0) \rangle }
\nonumber \\
 &=&  
\frac{ Z_P^0 (1+\tilde b_P a\tilde m_3/2) } 
	{ Z_A^0 \cdot Z_S^0 (1+ \tilde b_S a\tilde m_3/2) }\,,
	\label{ZPZS-1}
\end{eqnarray}
with $J = P$ or $A_4$.
Given $Z_A^0$, this also yields $ Z_P^0/Z_S^0 $.
Alternatively, one can use
\begin{eqnarray}
r_4 &\equiv &
\frac	{ \sum_{\vec{y}} e^{i\vec{p} \cdot \vec{y} }
	\langle \delta {\cal S}^{(12)}_I \ 
		P^{(23)}(\vec{y},y_4) \ J^{(31)}(0) \rangle }
	{ \sum_{\vec{y}}  e^{i\vec{p} \cdot \vec{y} }
	\langle S^{(13)}(\vec{y},y_4) \ J^{(31)}(0) \rangle }
\nonumber \\
 &=&  \frac{ Z_S^0 (1+\tilde b_S a \tilde m_3/2) } 
	{ Z_A^0 \cdot Z_P^0 (1+ \tilde b_P a\tilde m_3/2) }\,,
	\label{ZSZP-1} 
\end{eqnarray}
where for the source we choose either $ J^{(31)}= S^{(31)}$ or
$\sum_{\vec{z}} P^{(34)}(\vec{z},z_4) P^{(41)}(0) $ for $ 0 < y_4 < z_4$. 

%
%
The method described so far does not work for the tensor bilinear,
because the chiral rotation transforms it back into (other components of)
itself, and the dependence on $\tilde b_T$ cancels.
One can, however, use the method to determine $c_T$~\cite{rome-1}	.
For example, the AWI with ${\cal O} = T_{ij}$ at $\vec p=0$
and $J=T_{k4}$, can be rearranged into
\begin{eqnarray}
& & 1 + a c_T \frac{ \sum_{\vec{y}} \langle
    [- \partial_4 V_k ]^{(13)}(\vec{y},y_4) T_{k4}^{(31)}(0) \rangle }
    { \sum_{\vec{y}} \langle T_{k4}^{(13)} (\vec{y},y_4) T_{k4}^{(31)}(0) \rangle }
\nonumber \\
& & \hspace*{10mm} = Z_A^0 \frac { \sum_{\vec{y}} 
        \langle \delta {\cal S}^{(12)}_I \ T_{ij}^{(23)}(\vec{y},y_4) 
	\ T_{k4}^{(31)}(0) \rangle }
        { \sum_{\vec{y}}  
        \langle T_{k4}^{(13)} (\vec{y},y_4) \ T_{k4}^{(31)} (0) \rangle } \,,
        \label{cT-1}
\end{eqnarray}
Here we have moved the $c_T$ dependence in $(T_I)_{k4}$ onto the l.h.s.,
and used the fact that $(T_I)_{ij}$ has no contribution
from the $c_T$  term at $\vec p = 0$.
Given $Z_A^0$, this equation determines $c_T$.
A consistency check is that the result should be independent of $\tilde m_3$.

%
%
It turns out that one can determine $b_T$ using the AWI (\ref{AWI}), 
but to do so one must work with non-zero $m_1$ and $m_2$.
This means that the AWI is not completely improved: 
terms of $O(a)$ result from the contact of the pseudoscalar
density in $\delta S$ with the operator ${\cal O}$.
The key point, however, is that these terms are proportional to
$\tilde m_1+\tilde m_2$, while the dependence on $b_T$ is
proportional to the difference $\tilde m_1-\tilde m_2$.
By separating these two dependences one can, in principle, determine $b_T$.

To explain this in detail we recall that off-shell $O(a)$ improvement
requires the addition of an extra operator 
(multiplied by an extra improvement coefficient)
for each bilinear~\cite{rome-1}.
For the pseudoscalar and tensor the required additions are
\begin{eqnarray}
 P^{(12)} \longrightarrow P^{(12)} &+& 
    a c'_P \bar\psi^{(1)}  (-\overleftarrow{\slashnext D} + m_1) \gamma_5
	\psi^{(2)} \nonumber \\
    &+& a c'_P \bar\psi^{(1)} \gamma_5 (\overrightarrow{\slashnext D} + m_2) 
	\psi^{(2)} \,, 
\label{contactP} \\
 (T_I)_{\mu\nu}^{(23)} \longrightarrow (T_I)_{\mu\nu}^{(23)} &+& 
    a c'_T \bar\psi^{(2)}  (-\overleftarrow{\slashnext D} + m_2) 
	i\sigma_{\mu\nu} 
	\psi^{(3)} \nonumber \\
    &+& a c'_T \bar\psi^{(2)} i\sigma_{\mu\nu} (\overrightarrow{\slashnext D}
	 + m_3) 
	\psi^{(3)} \,.
\label{contactT}
\end{eqnarray}
Inserting the off-shell improved operators into the AWI considered
previously we find
\begin{eqnarray}
r_5 &\equiv& 
\frac { \sum_{\vec{y}} e^{i\vec{p} \cdot \vec{y} }
        \langle \delta {\cal S}^{(12)}_I \ (T_I)_{ij}^{(23)}(\vec{y},y_4) 
		\ T_{k4}^{(31)}(0) \rangle }
        { \sum_{\vec{y}} e^{i\vec{p} \cdot \vec{y} }
        \langle (T_I)_{k4}^{(13)} (\vec{y},y_4) \ T_{k4}^{(31)} (0) \rangle }
\nonumber \\  & = & 
	\frac{1 + a \tilde b_T (\tilde m_1 - \tilde m_2)/2} 
	{Z_A^0(1-a \tilde b_A\tilde m_{12})} - 2 a (c'_T + c'_P) \tilde m_{12}
	\,,        \label{bT-1}
\end{eqnarray}
Note that both the $\tilde b_A$ term and the contact terms
proportional to $c'_T+c'_P$ depend on $\tilde m_{12}$.  Thus, assuming
that $c_T$ is known, $\tilde b_T$ can be obtained from the dependence
of $r_5$ on $\tilde m_1 - \tilde m_2$. Note that the ratio $r_5$
depends on $c_T$ at both zero and non-zero momentum.

A similar extension can be considered for the other AWI discussed above.
It is straightforward to see that the $\tilde m_1-\tilde m_2$
dependence of the ratios $r_1$ and $r_2$ allows
a determination of $\tilde b_V+\tilde b_A$, 
while that of $r_3$ and $r_4$ gives a determination 
of $\tilde b_P+\tilde b_S$. Thus, in principle, one can determine
all the $\tilde b_X$ using the generic AWI.
One can also determine the five additional improvement constants
$c'_X$ using the dependence on $\tilde m_{12}$ as will be discussed in 
\cite{WIunquenched}.
Further consistency checks are provided by the
three-point vector WI with non-degenerate masses---although these
by themselves do not allow one to disentangle the $\tilde b_X$
from the $c'_X$.

\bigskip

Petronzio and di Divitiis
have shown that one can also use the two-point versions of the 
vector and axial WI to determine a subset of the quenched $b_X$,
namely $b_A-b_P$, $b_S$ and $b_V$~\cite{rome-3}.
The key point is again the use of non-degenerate quarks.
In our numerical study we use some of their results,
which we recall here.

%
%
The first result is obtained
by comparing the WI mass $\tilde m$ to the bare quark mass $m$.
These two masses are both related to the renormalized quark mass,
the former through eq.~(\ref{m(t)}), and the latter by
\begin{equation}
   m_i^R = Z_m^0 (1 + \tilde b_m a \tilde m_i)\ m_i 
\label{e:massH} 
\end{equation}
Combining these relations, and considering only degenerate masses $m_i=m_j$,
one finds
\begin{eqnarray}
\frac{ m } { \tilde m } &=& \frac{ Z_A^0}{Z_P^0 Z_m^0} 
[1 + (\tilde b_A - \tilde b_P - \tilde b_m) a\tilde m] 
  \label{mass2} \\
 &=& \frac{ Z_A^0 Z_S^{0} }{Z_P^0 } 
[1 + (\tilde b_A - \tilde b_P + \tilde b_S/2) a\tilde m] \,.
  \label{mass3}
\end{eqnarray} 
To obtain the second line 
we have used the results $ Z_m^0 Z_S^0 = 1 $ and $b_S = - 2 b_m$,
the latter valid only in quenched QCD \cite{LuscherbSbm}. 
Thus, from the slope and intercept of (\ref{mass3})
one can determine $\tilde b_A - \tilde b_P + \tilde b_S/2$ and
${ Z_A^0 Z_S^0 }/{Z_P^0 }$, respectively.
To use this method we need to determine $\kappa_c$, which
involves a chiral extrapolation.

%
%
The relationship between $\tilde m_{ij}$ and $(m_i+m_j)/2$
for non-degenerate quarks gives additional information.
It turns out that this information can be gleaned without reference
to the bare quark masses, and thus without the need for chiral
extrapolation. In particular, by enforcing
$(2m_1)^R + (2m_2)^R = 2(m_1+m_2)^R$ to $O(a)$, one finds
\begin{equation}
\tilde b_P - \tilde b_A = -
	\frac{ 4\tilde m_{12} - 2 [ \tilde m_{11} + \tilde m_{22} ]}
	{  a [ \tilde m_{11} - \tilde m_{22} ]^2 } \,.
\label{bP-bA}
\end{equation}
This result, in terms of bare masses, 
was already noted in Ref.~\cite{rome-3}.

The final result from Ref.~\cite{rome-3} uses the vector two-point WI.
Requiring $\partial_\mu (V_R)^{(12)}_\mu = (m_1^R-m_2^R) S_R^{(12)}$
leads to
\begin{eqnarray}
\Delta_{12} &\equiv&
\frac   { \sum_{\vec{x} }  e^{ i\vec{p} \cdot \vec{x} }
	\langle \partial_\mu {V_I}_\mu^{(12)} (\vec{x},t) J^{(21)}(0) \rangle }
        { \sum_{\vec{x} }  e^{ i\vec{p} \cdot \vec{x} }
	\langle S^{(12)}(\vec{x},t) J^{(21)}(0) \rangle }
\label{VS1}
\\
&=& 
\frac{Z_S [1 + \tilde b_S a \tilde m_{12}] } 
	{ Z_V [1 + \tilde b_V a \tilde m_{12}] } (m^R_1 - m^R_2)  \,.
\label{VS2}
\end{eqnarray}
We implement this using two sources: $ J^{(21)}= S^{(21)}$, and
$J^{(21)}=\sum_{\vec{z}} P^{(23)}(\vec{z},z_4) P^{(31)}(0)$ 
with $ 0 < t < z_4$. 
Enforcing the relation $(2m_1)^R - (2m_2)^R = 2(m_1-m_2)^R$ to $O(a)$, 
with the l.h.s. determined from the AWI (\ref{m(t)}) and
the rhs from the vector WI (\ref{VS1}), we find
\begin{eqnarray}
& & \frac{b_S - b_V}{2} +  (b_P - b_A ) 
       =  \frac{ \Delta_{12} - R_Z [ m_{11} - m_{22} ] }
	{ a R_Z [ m_{11}^2 - m_{22}^2 ]} \,,
\label{bS-bV}
\\
& & R_Z \equiv \frac{Z_S^0}{Z_P^0} \cdot \frac{Z_A^0}{Z_V^0} \,.
\end{eqnarray}
We can use eqs.~(\ref{bP-bA},\ref{bS-bV}) to determine
$b_P$ and  $b_S$ separately, 
since $ b_V $ and $b_A$ are already known from eqs. (\ref{ZV},\ref{cV1}).

\bigskip

We have performed a pilot test of our method on an ensemble of
$16^3\times 48$ quenched lattices at $\beta=6/g^2=6.0$.  We use the
tree-level tadpole-improved value for the clover coefficient, $c_{SW}
=1.4755$, rather than the non-perturbative value $c_{SW}=
1.769$~\cite{alpha-2}.  This implies that our results for the $b_X$
differ from the non-perturbative values by corrections of $O(g^2)$.
We do not have data with three non-degenerate quarks, and so can test
only the simpler version of our method which does not yield $b_T$.
Nevertheless, our results should suffice to assess the practicality of
using WI with non-degenerate quarks to determine the $b_X$
non-perturbatively.

Previous determinations of the improvement
coefficients $c_X$ have used Schr\"odinger functional
boundary conditions in the time direction, with sources $J$
placed on the boundaries. One advantage of
this approach is that one can work directly in the chiral limit.
In our study we use the same correlation functions as used in studying
the spectrum and decay constants, 
i.e. we have periodic boundary conditions in the time direction.
Thus a secondary output of our study is a comparison of
these two approaches for determining the $c_X$.
We stress, however, that our method for determining the $b_X$ works
for any choice of sources $J$, and in particular can be applied
to the Schr\"odinger functional.

The correlation functions required in the integrated axial Ward
identity (\ref{AWI}) are obtained as follows.  Quark propagators are
calculated using a Wuppertal smeared source at $t=0$ for five
different values of $\kappa$ corresponding to $aM_\pi = 0.57, 0.50,
0.43, 0.35, 0.24$.  These propagators are used both to construct
two-point functions and also as sources for propagators with the
insertion of $\delta S_I$ defined in eq.~(\ref{deltaS12}).  Our
insertion volume ${\cal V}$ is the region between $t=4$ and $18$.  The
second inversion uses the same $\kappa$, so that, as already noted,
$m_1 = m_2$.  To construct three-point functions, propagators with and
without sources are contracted to form the operator ${\cal O}$.  This
allows us to insert any momentum into ${\cal O}$ and to place it
anywhere in the interval $4 < t < 18$. 

For the chiral extrapolations we ignore the correlations in the data
between the different mass points. We find poor
signals in some of the correlators containing quarks of the
lightest mass, and so exclude the lightest mass from
the extrapolations. The remaining four values of quark mass
correspond to the range $m_s - 2.6m_s$, where $m_s$ is the physical
strange quark mass. Note that the relevant expansion parameter here
is $m_q a$, and this is small for our range of quark masses.

Our results from the various WI are summarized in
Table~\ref{t:Zresults}.  Where there is a choice of sources, $J$, we
have quoted the results with the best signal.  In the following we
discuss the various determinations pointing out salient features.

%
\begin{table}[tb]
\setlength{\tabcolsep}{1pt}
\renewcommand{\arraystretch}{1.2}
\begin{center}
\begin{tabular}{|l|l|l|l|}
\hline\thinspace
  eq. 		& \thinspace observable 	
			& \thinspace two-point $\partial$ & 	\thinspace three-point $\partial$	\\ 
\hline\thinspace
 \ref{cA}	&\thinspace $ c_A $					&\thinspace $ -0.005 (   15 )  $ &\thinspace $ -0.025 (   10 )  $ \\\thinspace
 \ref{ZV}    	&\thinspace $ Z^0_V $					&\thinspace $ +0.745 (    1 )  $ &\thinspace $ +0.746 (   1 )  $ \\\thinspace
 \ref{ZV}    	&\thinspace $\tilde b_V$				&\thinspace $ +1.57  (   2  )  $ &\thinspace $ +1.57  (   2 )  $ \\\thinspace
 \ref{cV1}   	&\thinspace $ Z^0_V $					&\thinspace $ +0.753 (    6 )  $ &\thinspace $ +0.759 (   6 )  $ \\\thinspace
 \ref{cV1} 	&\thinspace $ \tilde b_A - \tilde b_V $ 		&\thinspace $ -0.56 (   9  )  $ &\thinspace $ -0.48 (  10 )  $ \\\thinspace
 \ref{cV2}i 	&\thinspace $ c_V $              			&\thinspace $ -0.66 (  27 )  $ &\thinspace $ -0.65 (  25 )  $ \\\thinspace
 \ref{ZAZV-1}i	&\thinspace $ Z^0_V/(Z^0_A)^2 $ 			&\thinspace $ +1.41 (  12 )  $ &\thinspace $ +1.38 (  11 )  $ \\\thinspace
 \ref{ZAZV-1}i	&\thinspace $\tilde b_A - \tilde b_V $			&\thinspace $ +1.2 ( 1.2 )  $ &\thinspace $ +1.8 ( 1.0 )  $ \\\thinspace
 \ref{ZAZV-1}ii &\thinspace $ c_V  $					&\thinspace $ -0.30 (  26 )  $ &\thinspace $ -0.31 (  24 ) 
 $ \\\thinspace
 \ref{ZAZV-2}	&\thinspace $ Z^0_A $               			&\thinspace $ +0.76 (   2 )  $ &\thinspace $ +0.77 (   2 )  $ \\\thinspace
 \ref{ZAZV-3}	&\thinspace $ Z^0_A/Z^0_V $				&\thinspace $ +0.97 (   5 )  $ &\thinspace $ +0.98 (   4 )  $ \\\thinspace
 \ref{ZAZV-3}	&\thinspace $\tilde b_A\!-\!\tilde b_V $		&\thinspace $ +0.38 (  74 )  $ &\thinspace $ +1.09 (  69 )  $ \\\thinspace
 \ref{ZPZS-1}	&\thinspace $Z^0_A Z^0_S/Z^0_P$     			&\thinspace $ +0.95 (   2 )  $ &\thinspace $ +0.98 (   1 )  $ \\\thinspace
 \ref{ZPZS-1}i	&\thinspace $\tilde b_P\!-\!\tilde b_S $  		&\thinspace $ -0.09 (   9 )  $ &\thinspace $ -0.06 (   9 )  $ \\\thinspace
 \ref{cT-1}	&\thinspace $ c_T $					&\thinspace $ +0.17 (   7 )  $ &\thinspace $ +0.18 (   6 )  $ \\\thinspace
 \ref{mass3}  	&\thinspace $ Z^0_A Z^0_S/Z^0_P $   			&\thinspace $ +0.97 (   1 )  $ &\thinspace $ +0.97 (   1 )  $ \\\thinspace
 \ref{mass3}	&\thinspace $\tilde b_A\!-\!\tilde b_P+\tilde b_S/2 $	&\thinspace $ +0.48 (   1 )  $ &\thinspace $ +0.50 (   1 )  $ \\\thinspace
 \ref{bP-bA}	&\thinspace $\tilde b_P \!-\!\tilde b_A $           	&\thinspace $ -0.20 (   9 )  $ &\thinspace $ -0.13 (   8 )  $ \\\thinspace
 \ref{bS-bV}	&\thinspace $\tilde b_S\!-\!\tilde b_V\!-\!2\tilde b_P\!-\!2\tilde b_A $  &\thinspace $ +0.04 (  48 )  $ &\thinspace $ -0.27 (  38 )  $ \\
\hline
\end{tabular}
\vspace{10pt}
\caption{Results for improvement coefficients from the listed WI.
The two columns of results are for two choices of
discretization of derivatives in eq.~(\ref{cA}),
as discussed in the text.
The labels ``i''--``v'' are explained in the text. 
}
\label{t:Zresults}
\end{center}
\end{table}
%

We determine $c_A$ by requiring that the right hand side of
eq.~(\ref{cA}) is as close to a constant as possible over a range
of times $t_{\rm min}-t_{\rm max}$. For long times only the pion
contributes and the ratio is constant for all values of $c_A$---thus 
any choice of $t_{\rm max}$ in the pion-dominated region is equally good.
To maximise our sensitivity to $c_A$, we choose $t_{\rm min}$ as small 
as possible while avoiding contact between $\delta S_I$ and $J$.
We find that our results are insensitive to small 
variations in $t_{\rm min}$.
There is a similar insensitivity to the choice of source $J$.
This is in marked contrast to results from the unimproved action 
($c_{SW}= 0$) \cite{lat98us-2}
and gives us confidence that our tadpole-improved action has small
enough $O(g^2 a)$ errors that we can carry out our tests.

To estimate the size of $O(a^2)$ errors (which become
$O(a)$ errors in $c_A$), 
we use two choices of discrete derivatives in eq.~(\ref{cA}):
a scheme based on two-point derivatives,
\begin{eqnarray*}
\partial f(x\!+\!1/2) &=& f(x\!+\!1)-f(x)\,, \\
\partial^2 f(x\!+\!1/2) &=& {f(x\!+\!2)-f(x\!+\!1)-f(x)\!+\!f(x-1) \over2}
 \,,\\
f(x\!+\!1/2) &=& {f(x\!+\!1)+ f(x) \over 2}\,,
\end{eqnarray*}
and one based on three-point derivatives,
\begin{eqnarray*}
\partial f(x) &=& (f(x+1)-f(x-1))/2\,,\\ 
\partial^2 f(x) &=& f(x+1)-2f(x)+f(x-1)\,.
\end{eqnarray*}
The $O(a^2)$ correction in the three-point $\partial f$ is four times
larger than in the two-point case.  The results for $c_A$ from these two
discretization schemes differ at $O(a)$.\footnote{Since our action
is only perturbatively improved, one might worry that the two
determinations of $c_A$ could differ by $O(g^2)$ in addition to
$O(a)$. This is not the case, however, because we use essentially the
same matrix elements in both determinations, and so the difference
between them is explicitly proportional to $a$.}  As shown in
Table~\ref{t:Zresults}, this difference is substantial as,
unfortunately, are the statistical errors.  Because of this
difference, we present results for the remaining improvement constants
using both choices of $c_A$.

Both discretization schemes yield results for $c_A$ significantly
different from the value $c_A=-0.083(7)$ obtained previously at
$c_{SW}= 1.769$ using the Schr\"odinger functional~\cite{alpha-2}.
This difference is presumably a combination of $O(g^2)$ and $O(a)$
effects. We are unable to resolve this discrepancy in this work.


%
The signal in the vector WI (\ref{ZV}) is very good, and provides our
best estimates of $Z_V$ and $b_V$.  There is a tiny dependence on $c_A$
due to the determination of the quark mass using eq.~\ref{cA}. 

\begin{figure}[tb]
\begin{center}
\epsfxsize=0.9\hsize 
\epsfbox{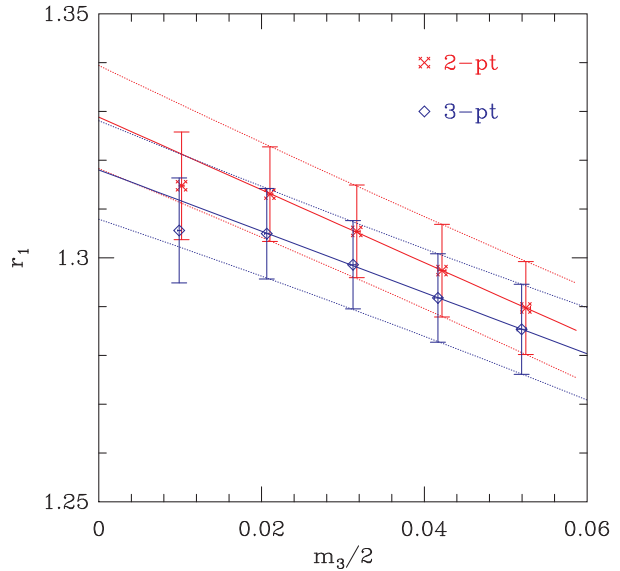}
\vspace*{-10mm}
\end{center}
\caption{The ratio $r_1$ as a function of $\tilde m_3/2$,
after linear extrapolation to $\tilde m_1=\tilde m_2=0$. 
The intercept and slope give $1/Z_V^0$ and 
$\tilde b_A-\tilde b_V$, respectively. The 
fit excludes the lightest point as the 
expected plateau in the fits to the different $r_i$ are not 
uniformly good. }
\label{fig:cV1}
\end{figure}

We now turn to results from our new method.  We use a two-point
discretization of the derivative in $\delta S_I$
throughout.\footnote{The results for improvement constants are
expected to be fairly insensitive to this choice of discretization
because, after integration over ${\cal V}$, different choices differ
only by surface terms.}  In Fig.~\ref{fig:cV1} we show our results for
the quantity $r_1$ which appears in the AWI (\ref{cV1}).  The data
show a linear dependence on $\tilde m_3$ for our range of masses. 

There is a numerical subtlety in the extraction of 
$\tilde b_A -\tilde b_V$.  
When using eq.~\ref{cV1}, we have to make two choices concerning the 
form of $A_I^{(13)}$:
whether to discretize it using two- or three-point derivatives;
and whether to use the $m$ dependent value of $c_A$
obtained from eq.~\ref{cA}, or the chirally extrapolated value.
Both choices only effect the result for $\tilde b_A - \tilde b_V$ at $O(a)$.
In particular, it is straightforward to see that the two options for $c_A$
lead to results for $\tilde b_A - \tilde b_V$ differing by $\sim a M_\pi^2/m$
(assuming pion domination of the correlators).
Since $M_\pi^2/m$ is a large scale, perhaps as large as 5 GeV, these
differences, although technically of higher order, 
can be numerically significant.
We find that they are only 
a 15\% effect for the two-point discretization of $A_I$,
but are much larger for the three-point discretization.
For this reason we use the two-point discretization.
As for the choice of $c_A$, we take the chirally extrapolated value,
since this is the consistent choice at the order of improvement we are working.
This does, however, have the disadvantage of giving poorer plateaus
in the ratio $r_1$.\footnote{We have a similar choice
for the $c_A$ appearing in $\delta S_I^{(12)}$, and here we use
the mass dependent value which leads to better signals. 
This does not, however, directly affect our results for the $\tilde b_X$
since we extrapolate to $\tilde m_{12}=0$.}

The results for $Z_V^0$ and $\tilde b_A-\tilde b_V$ are given in
Table~\ref{t:Zresults}. $Z_V^0$ is consistent, at the two $\sigma$
level, with the result from the vector WI, but has much larger errors.
$\tilde b_A-\tilde b_V$ is determined with a $\sim 20\%$ error.

The determination of $c_V$ is more problematic and we have attempted
a number of approaches. The two best are\looseness-1
\begin{itemize}
\item[(i)] We obtain $c_V$ by first equating the right hand sides of
Eqs.~(\ref{cV1},\ref{cV2}) for each $\tilde m_3$ and then extrapolating 
to $\tilde m_3=0$;
\item[(ii)] 
For each $\tilde m_1= \tilde m_2$ and $\tilde m_3$, we solve 
$r_1 r_2 = r_3 r_4$, where only $r_2$ depends on $c_V$. The chiral 
extrapolation in $\tilde m_1= \tilde m_2$ removes the contribution of 
the contact terms. Lastly we extrapolate to $\tilde m_3=0$. 
\end{itemize}
All methods require knowledge of $c_A$.
The correction term proportional to $c_V$ is 
$\approx 4\%$ in eq.~(\ref{cV2})
and $\approx 20\%$ in eq.~(\ref{ZAZV-1}), so one
needs high statistics to get an accurate estimate.
We prefer (i) as it is the most direct, and because $r_4$ does not 
have a good plateau. The large uncertainty in $c_V$ extracted by these
methods accounts for $\sim 50\%$ of the errors in $Z_A^0$, and
subsequently in $Z_P^0 / Z_S^0$ and $c_T$.

With $c_V$ (and $Z_V^0$) in hand, we use eq.~(\ref{ZAZV-2}) to
determine our best estimate of $Z_A^0$. The results from
eqs.~(\ref{ZAZV-1}) and (\ref{ZAZV-3}) are consistent but have
larger errors. The estimates of $\tilde b_A-\tilde b_V$ from 
these two equations have very large errors. 

\begin{figure}[tb]
\begin{center}
\epsfxsize=1.0\hsize 
\epsfbox{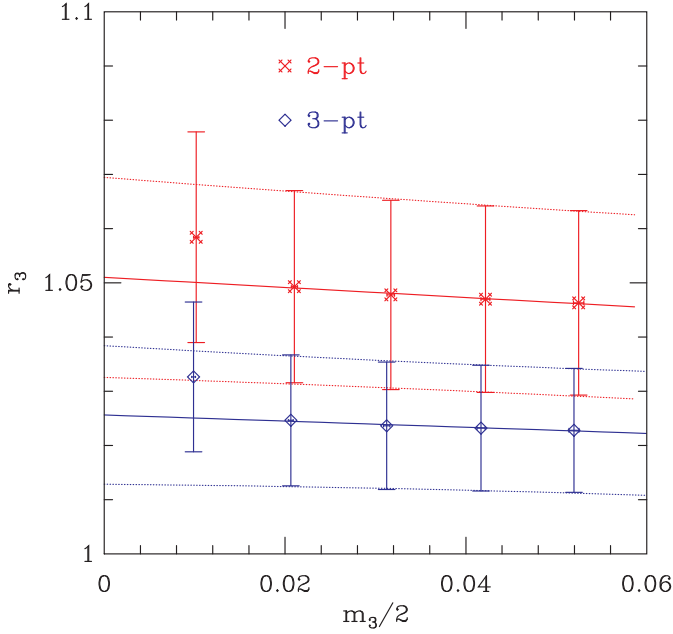}
\end{center}
\vspace*{-10mm}
\caption{The ratio $r_3$ as a function of $\tilde m_3/2$,
after linear extrapolation to $\tilde m_1=\tilde m_2=0$. 
The intercept and slope give $Z_P^0/(Z_S^0 Z_A^0)$ and 
$\tilde b_P-\tilde b_S$, respectively. 
The fit excludes the lightest point.}
\label{fig:r3}
\end{figure}

The final application of our new method is the determination
of $\tilde b_P-\tilde b_S$ using eqs.~(\ref{ZPZS-1}) and (\ref{ZSZP-1}).
Note that neither of these requires knowledge of $c_V$.
We find a good signal in the ratio $r_3$ but not in $r_4$.
The former is shown in Fig.~\ref{fig:r3} and the slope yields the
value quoted in the table.
The intercept gives an estimate of $Z_P^0/Z_S^0 Z_A^0$.

The last mixing coefficient $c_T$ is determined from eq.~(\ref{cT-1}).
There is a good signal in all correlation functions and only $c_A$
and $Z_A^0$ are needed beforehand. The error is dominated by the
uncertainty in $c_V$ which feeds in through $Z_A^0$.

Further information, and consistency checks on the previous results,
are provided by the WI using two-point functions.  Equation
(\ref{mass3}) gives a result for $Z_P^0/Z_S^0 Z_A^0$, consistent with
that from above, with similar errors.  It also gives a very accurate
result for the combination $\tilde b_A-\tilde b_P + \tilde b_S/2$,
with little dependence on $c_A$.  There is, however, an additional
uncertainty due to the choice of $\kappa_c$.  
%
The two remaining equations, (\ref{bP-bA}) and
(\ref{bS-bV}), which do not require $\kappa_c$, give rather poor
determinations.  An important technical point is that the
$O(a)$ correction in eq.~(\ref{VS1}) is large when the source
$J^{(21)}$ is chosen to be $S^{(21)}$ as $M_{scalar} a \approx 1$. 
In order to get results
consistent with those from the choice $J^{(21)} = \sum_{\vec z}
P^{(23)}(\vec z, z_4) P^{(31)}(0)$, we needed to use a five-point
discretization of $\partial_\mu V_\mu$ when using the source
$J^{(21)}=S^{(21)}$.

\begin{table}[tb]
\setlength{\tabcolsep}{1pt}
\begin{center}
\begin{tabular}{|c|l|c|c|}
\hline                        		&\thinspace  LANL                &\thinspace  ALPHA       &\thinspace Pert. Th. \\
 & & & \\[-9pt]
\hline
 & & & \\[-9pt]
 $ C_{SW} $             	 	&\thinspace  1.4755              &\thinspace  1.769       &\thinspace$+1$        \\
 & & & \\[-9pt]
\hline 
 & & & \\[-9pt]
 $ Z^0_V $              		&\thinspace$ +0.745 (1)(1) $  &\thinspace  0.7809(6)   &\thinspace$+0.810$ \\\thinspace
 $ Z^0_A $              		&\thinspace$ +0.76  (2)(1) $  &\thinspace  0.7906(94)  &\thinspace$+0.829$ \\\thinspace
 $ Z^0_P/Z^0_S $        		&\thinspace$ +0.77  (4)(1) $  &\thinspace  N.A.        &\thinspace$+0.949$ \\
 & & & \\[-9pt]
 \hline
 & & & \\[-9pt]
 $ c_A $                		&\thinspace$ -0.005 (15)(20) $ \negthinspace\vrule width 0pt  &\thinspace  $-0.083(5)$ &\thinspace$-0.013$ \\\thinspace
 $ c_V $                		&\thinspace$ -0.66  (27)(02) $  &\thinspace  $-0.32(7)$  &\thinspace$-0.028$ \\\thinspace
 $ c_T $                		&\thinspace$ +0.17  (7)(1)   $  &\thinspace  N.A.        &\thinspace$+0.020$ \\
 & & & \\[-9pt]
 \hline                                                            
 & & & \\[-9pt]
 $\tilde b_V $          		&\thinspace$ +1.57 (2)(1) $  &\thinspace  N.A.        &\thinspace$+1.106$ \\\thinspace
 $ b_V $                		&\thinspace$ +1.62 (3)(1) $  &\thinspace  1.54(2)     &\thinspace$+1.273$ \\\thinspace
 $\tilde b_A-\tilde b_V$		&\thinspace$ -0.56 (9)(8) $  &\thinspace  N.A.        &\thinspace$-0.002$ \\\thinspace
 $\tilde b_P-\tilde b_S$		&\thinspace$ -0.09 (9)(3) $  &\thinspace  N.A.        &\thinspace$-0.066$ \\\thinspace
 $\tilde b_P-\tilde b_A$		&\thinspace$ -0.20 (9)(7) $  &\thinspace  N.A.        &\thinspace$+0.002$ \\\thinspace
 $\tilde b_A-\tilde b_P+\tilde b_S/2$	&\thinspace$ +0.48 (1)(2) $  &\thinspace  N.A.        &\thinspace$+0.584$\\\thinspace
 $ b_A- b_P+ b_S/2$			&\thinspace$ +0.49 (2)(2) $  &\thinspace  N.A.        &\thinspace$+0.673$\\\thinspace
 & & & \\[-9pt]
 \hline
 & & & \\[-9pt]
 $\tilde b_A$                  &\thinspace$ +1.01 (  09 )(09)  $ &\thinspace  N.A.  & \thinspace$+1.104$    \\\thinspace
 $\tilde b_P$                  &\thinspace$ +0.81 (  14 )(16)  $ &\thinspace  N.A.  & \thinspace$+1.105$     \\\thinspace
 $\tilde b_S$                  &\thinspace$ +0.90 (  17 )(13)  $ &\thinspace  N.A.  & \thinspace$+1.172$    \\[3pt]
\hline
\end{tabular}
\vspace{10pt}
\caption{Our best estimates for normalization and improvement coefficients,
compared to previous results where available, and to
tadpole improved 1-loop perturbation theory (using $u_0=0.8778$). 
An estimate of the perturbative errors is $\alpha_s^2 \sim 0.02$. 
The first error is statistical, the second an estimate of
the $O(a^2)$ error. See text for details.}
\label{t:Zcomparison}
\end{center}
\end{table}

We collect our best results in Table~\ref{t:Zcomparison}.  These are
obtained using the two-point derivative in $\delta S_I$. The
difference between two- and three-point discretization is added as an
additional error. Our results are compared to previous
non-perturbative results, and to those of tadpole-improved 1-loop
perturbation theory. The latter are obtained using 1-loop results
available in the literature
\cite{martinelli91,sint97-1,sint97-2,jlqcd98}.  We note that for
the $c_X$ and the $\tilde b_X$ tadpole improvement is equivalent to
using the boosted coupling $g^2/u_0^4$ in the 1-loop result.  The same
is not true for the $Z_X$ and $b_X$.  
The conversion factor between the two definitions of the
mass-dependent improvement coefficients, $\tilde b_X/b_X=Z_A^0
Z_S^0/Z_P^0$, turns out to be very close to unity (see
Table~\ref{t:Zresults}).  Thus, with the exception of $b_V$ and
$b_A-b_P+b_S/2$, which are determined quite accurately, we do not
convert our results back to the standard definition $b_X$.



We draw the following conclusions.  First, the method we have
introduced appears to have practical utility.  In the best channels,
the statistical and systematic errors in the determination of the
differences $\tilde b_A - \tilde b_V$ and $\tilde b_P - \tilde b_S$
are small compared to the values, $\tilde b_X\approx 1$, of the
coefficients themselves.
Second, $c_V$ is determined rather poorly using our WI, and this
accounts for a substantial fraction of the errors in $Z_A^0$, $Z_P^0 /
Z_S^0$, and $c_T$.  For $c_A$ and $c_V$, the Schr\"odinger functional
method~\cite{alpha-1,alpha-2,alpha-3,alpha-4} gives results with much
smaller errors. 
Third, we have found, in some cases, substantial disagreement between
the perturbative predictions and our non-perturbative results.  The
most striking cases are $b_V$ and $b_A - b_V$.  These differences
could be effects of $O(a^2)$,
and we intend to investigate this issue by repeating the calculations
at different values of $a$.  
Finally, there are also differences between our
results and those of the ALPHA Collaboration.  We anticipate that these
differences are due in part to our use of an action that is not fully
$O(a)$ improved. We have initiated simulations at $c_{SW}=1.769$ to
verify this.

%
%
We acknowledge the support of the Advanced Computing Laboratory at Los Alamos.
This work was supported in part by
the U.S. Department of Energy grants DE-LANL-ERWE161, DE-FG02-96ER40945,
and DE-FG03-96ER40956. 
SS is very grateful to the Center for Computational Physics at the University
of Tsukuba for the hospitality received there while completing this work.

\end{document}